\documentclass[aps, 
               prb, 
               twocolumn, 
               10pt, 
               superscriptaddress,
               amsfonts, 
               amsmath, 
               amssymb]{revtex4-2}

\usepackage{bm}
\usepackage{braket}

\usepackage[version=4]{mhchem}
\usepackage{microtype}
\usepackage{upgreek}
\usepackage{graphicx}
\usepackage{hyperref}
\usepackage{xcolor}

\hypersetup{
    colorlinks,
    linkcolor={blue!80!black},
    citecolor={blue!80!black},
    urlcolor={blue!80!black}
}

\usepackage{xcolor}

\let\Im\relax

\DeclareMathOperator{\Im}{Im}

\DeclareMathOperator{\tr}{tr}
\newcommand{\dd}[0]{\mathrm{d}}

\begin{document}
\title{Impurity states in altermagnetic superconductors}

\author{Andrea Maiani}
\affiliation{Nordita, KTH Royal Institute of Technology and Stockholm University,
Hannes Alfvéns väg 12, SE-10691 Stockholm, Sweden}
\author{Rubén Seoane Souto}
\affiliation{Instituto de Ciencia de Materiales de Madrid (ICMM), Consejo Superior de Investigaciones Científicas (CSIC), Sor Juana Inés de la Cruz 3, E-28049 Madrid, Spain}

\date{\today}

\begin{abstract}
Altermagnetic superconductors offer the possibility of exploring unconventional superconductivity, including topological states and finite-momentum superconductivity, with promising applications in spintronics and quantum information. However, a direct experimental confirmation of their existence remains elusive. In this work, we propose non-magnetic impurities as probes of the interplay between altermagnetism and superconductivity. These impurities induce spin-polarized subgap states whose spatial extension reflects the magnetic order of the substrate material. Depending on whether or not the impurity respects the bulk symmetries, these states form spin-degenerate or spin-split doublets. An external magnetic field aligned with the Néel vector can further control the doublet splitting. We further demonstrate that coupling between impurity states leads to a position-dependent, spin-sensitive hybridization, enabling another approach for \textit{in situ} control of atomic-size quantum devices. These findings provide unambiguous experimental signatures of altermagnetic superconductivity accessible via local measurements such as scanning tunneling microscopy and open unexplored pathways for designing tunable quantum devices.
\end{abstract}
\maketitle

\section{Introduction}

Recently, a new form of magnetic order has been discovered in nature: Altermagnets feature a vanishing net magnetization caused by the presence of sublattices with opposite spins that are connected by a crystal symmetry other than real-space translation or inversion~\cite{Naka_2019_Spin, Naka_2020_Anomalous, Smejkal_2020_Crystal, Mazin_2022_Editorial, Smejkal_2022_Emerging, Olsen_2024_Antiferromagnetism,  Autieri_2024_New}. 
Unlike other magnetic materials, altermagnets feature multipolar order which separately breaks time-reversal and crystal rotation symmetries, but preserves their combinations leading to spin-split bands at finite momentum, with zero net magnetization, and unlifted spin degeneracy at zero momentum. Signatures of altermagnetism have been found in various types of materials such as metallic oxide \ce{RuO2}~\cite{Ahn_2019_Antiferromagnetism, Smejkal_2022_Emerging}, epitaxial films of \ce{Mn5Si3}~\cite{Reichlova_2020_Observation}, semiconducting cuprate \ce{La2CuO4}~\cite{LopezMoreno_2016_First, Smejkal_2022_Conventional}, \ce{\alpha-MnTe}~\cite{Lee_2024_Broken, Osumi_2024_Observation, Krempasky_2024_Altermagnetic, Kazmin_2024_Andreev}, perovskite \ce{CaMnO3}~\cite{Smejkal_2022_Conventional}, chalcogenide \ce{CoNb3S6}~\cite{Smejkal_2022_Conventional}, and several other materials~\cite{Fernandes_2024_Topological, Reimers_2024_Direct}.

The interplay between altermagnetism and superconductivity has attracted interest because of the resemblance between the altermagnet band structure and the $d$-wave superconducting pairing. Indeed, the peculiar multipolar order makes this material class interesting for engineering new materials with exotic properties~\cite{Hayami_2019_Momentum}. This has spurred great interest in understanding the competition between the two orders~\cite{Zhang_2024_Finite, Zyuzin_2024_Magnetoelectric, Ouassou_2023_dc, Sun_2023_Andreev, Zhu_2023_Topological, Chakraborty_2024_Zero, Wei_2023_Gapless, Giil_2024_Superconductor}, particularly in Josephson junctions~\cite{Beenakker_2023_Phase, Ouassou_2023_dc, Lu_2024_phi}, and their connection to topological states~\cite{Li_2023_Majorana, Li_2024_Realizing, Ghorashi_2023_Altermagnetic, Zhu_2024_Field}. The combination of altermagnetism and superconductivity has been predicted to occur naturally in some materials  (see, for example, Refs.~\cite{Smejkal_2022_Conventional, Brekke_2023_Two}). Alternatively, it can be engineered in heterostructures combining altermagnets and superconductors. Independently of the platform, a demonstration of the coexistence of the two effects is still lacking.

In this work, we propose non-magnetic (spinless) impurities as probes of altermagnetic superconductors' properties. In conventional superconductors, potential impurities typically do not induce subgap states~\cite{Anderson_1959_Theory, Balatsky_2006_Impurity}. However, in the presence of classical spin impurities, localized spin-polarized subgap states, known as Yu-Shiba-Rusinov (YSR) states, can emerge~\cite{Yu_1965_Bound, Shiba_1968_Classical, Rusinov_1969_Theory}. Notably, a material with the combination of superconductivity and altermagnetism may feature at the same time an open, unpolarized spectral gap, and a magnetic order, offering the necessary ingredients for the appearance of localized bound states. An impurity acts as an electron scatterer, revealing the finite-momentum properties of electrons in real space through the emergence of subgap states with distinct spatial profiles for their spin components. The corresponding problem without superconductivity has been recently studied in the metallic phase, where impurities lead to Friedel's oscillations~\cite{Chen_2024_Impurity, Sukhachov_2024_Impurity}. In superconductors, a strong impurity potential can cause a local $\pi$ shift resembling the physics of YSR states~\cite{Balatsky_2006_Impurity, Meng_2015_Superconducting}, opening the door to applications for topological superconductivity~\cite{Flensberg_NRM2021, Schrade_2015_Proximity}.

We finally analyze the coupling between subgap states at two potential impurities. The peculiar spin structure of the subgap states favors an orientation-dependent hybridization of the subgap states, reducing the energy of subgap states with parallel spins. The spin of the lowest-energy state depends on the relative angle between the impurities with respect to the lattice, a unique feature due to altermagnetism, absent in conventional superconductors. An external magnetic field can control the energy of the subgap states, eventually making the opposite-spin bound state the lowest in energy, characterized by a low hybridization between the impurity states. This magnetic control can be used to tune the hybridization between spin-polarized states formed close to impurities on superconductors. Spatially resolved measurements, such as scanning tunneling microscopy, are suitable to study the spatial extension of the subgap states, being proof of altermagnetism in superconductors.

\section{Model}

We consider a two-dimensional electron gas featuring $d$-wave altermagnetism, $s$-wave superconductivity, and Rashba spin-orbit coupling using a tight-binding model on a square lattice with constant $a$. At the mean-field level, the model can be expressed by the Hamiltonian
$   
    H = \frac{1}{2} \psi^\dag \mathcal{H} \psi + \frac{|\Delta_0|^2}{2 g}\,,
$
where $\mathcal{H}$ is the Bogoliubov--de Gennes Hamiltonian in the time-reversed hole basis $\psi^\dag = (\psi^\dag_{\uparrow}, \psi^\dag_{\downarrow}, -\psi_{\downarrow}, \psi_{\uparrow})$, $g$ is the two-body coupling strength, and $\Delta_0$ is the singlet pairing potential. We consider $\mathcal{H}$ as the sum of a bulk and impurity term,
$
    \mathcal{H} = \mathcal{H}_\mathrm{b} + \mathcal{H}_\mathrm{imp}
$.
The bulk term reads as
\begin{align}
\begin{split}
\label{eq:H_bulk}
    \mathcal{H}_\mathrm{b} = &[-t (L_y + L_x) - E_\mathrm{F}] \sigma_0\tau_z + \Delta_0' \sigma_0 \tau_x - \Delta_0'' \sigma_0 \tau_y \\
    + &[-t_\mathrm{so} i (D_y \sigma_x - D_x \sigma_y)] \tau_z  \\
    + &[- t_\mathrm{am} (L_y - L_x) \bm{n} +  \bm{b} ] \cdot  \bm{\sigma}  \tau_0  \\
\end{split}
\end{align}
where $\tau_j$ and $\sigma_j$ are the Pauli matrices in spin and Nambu subspaces. Here, $t$ is the hopping amplitude, $E_\mathrm{F}$ is the chemical potential measured from the bottom of the band, $t_\mathrm{am}$ is the altermagnetic hopping amplitude with $\bm{n} = \bm{e}_z$ being the Néel vector, $\Delta_0'={\rm Re}(\Delta_0)$, $\Delta_0''={\rm Im}(\Delta_0)$, $t_\mathrm{so}$ is the Rashba spin-orbit coupling strength, and $\bm{b}$ is an externally applied Zeeman field. 
The operators $D_x$ and $D_y$ represent discrete first-order derivatives along the $x$ and $y$ directions, respectively, defined as  
$
D_x \ket{i,\ j} = \frac{1}{2} \left( \ket{i+1,\ j} - \ket{i-1,\ j} \right),
$
and analogously for $D_y$. The operators $L_x$ and $L_y$ correspond to the discrete Laplacian along $x$ and $y$, respectively, defined as  
$
L_x \ket{i,\ j} = \ket{i+1,\ j} - 2 \ket{i,\ j} + \ket{i-1,\ j},
$
and similarly for $L_y$.
The altermagnetic term can be seen as an unconventional $d_{x^2-y^2}$-wave Zeeman field due to the exchange-dependent hopping. Physically, we can understand this term as an exchange field induced in the itinerant electrons when hopping. The exchange-dependent hopping is opposite between the x and y axes, with a direction determined by $\bm{n}$. The strength of the altermagnetism can be identified by the nondimensional parameter $\eta_\mathrm{am} = t_\mathrm{am}/t$.

The bulk Hamiltonian breaks both the time-reversal symmetry $\mathcal{T} = - i \sigma_y \mathcal{K}$, with $\mathcal{K}$ being complex conjugation, and the space symmetries in the point group D4. However, it preserves the symmetries generated by their combinations $\{ \mathcal{T} \mathcal{C}_{4z}\,, \mathcal{T} \mathcal{M}_{xy} \}$ where $\mathcal{C}_{4z}$ is the four-fold rotational symmetry and acts as $\mathcal{C}_{4z} |k_x\, k_y\, \sigma\rangle = |-k_y\, k_x\, \sigma\rangle $, while $\mathcal{M}_{xy}$ is the mirror operator and acts as $\mathcal{M}_{xy} |k_x\, k_y\, \sigma\rangle = |k_y\, k_x\, \sigma\rangle$. Note that, $\mathcal{T} \mathcal{C}_{4z}$ and $\mathcal{T} \mathcal{M}_{xy}$ can be interpreted as generalized time-reversal symmetry operators since they are antiunitary, and square to $-1$. Therefore, the degeneracy is protected by a weaker version of Kramers' theorem that relies on also preserving the crystal symmetries. The addition of Rashba spin-orbit coupling breaks the inversion symmetry but leaves the spin-degeneracy unlifted in real space. In this work, we assume the external magnetic field to be sufficiently weak so we can neglect orbital effects in the superconductor.

We simulate the impurities' effect as a renormalization of the chemical potential around the impurity site with a Hamiltonian term 
\begin{equation}
    \mathcal{H}_\mathrm{imp}(\bm{r}_{ij}) = \sum_\nu f(\bm{r}_{ij} - \bm{r}_\nu) V_0 \sigma_0 \tau_z\,, 
\end{equation}
where $f(\bm{r})$ is a function that describes the impurities' shape, $V_0$ is a parameter controlling the potential strength, $\bm{r}_\nu$ are the positions of the impurities, labeled by $\nu$, and $\bm{r}_{ij} = a\ (i,\ j)$ are the lattice site positions. For isotropic impurities, we will parametrize as a Gaussian curve $f(\bm{r}) = \exp(-r^2/2 w_\mathrm{imp}^2)$, although the main findings do not depend on this choice. 


Before considering the effect of the impurities, we analyze the density of states of the bulk Hamiltonian of a pristine altermagnetic superconductor. The effect of the altermagnetic order for itinerant electrons can be described as a momentum-dependent exchange field $\bm{h}(\bm{k})$. For the square lattice with $d$-wave altermagnetism, this field takes the form $\bm{h}(\bm{k}) = t_\mathrm{am} \left[\cos(k_y) - \cos(k_x)\right] \bm{n}$~\cite{Smejkal_2022_Giant, Smejkal_2022_Emerging}. The effective exchange field causes a direction-dependent spin-splitting similar to a Zeeman field.
It satisfies the $\mathcal{C}_4\mathcal{T}$ symmetry, vanishes at points where $k_x = \pm k_y$, including $\Gamma$ and $M$ points, while it reaches its maximum at the $X$ and $Y$ points.

\begin{figure}[ht]%
\centering
\includegraphics{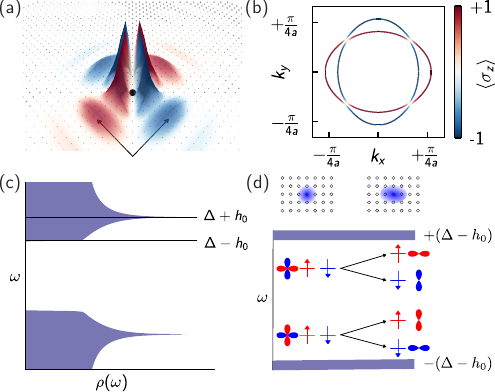}
\caption{
\label{fig1}
(a) Sketch of subgap states formed around an impurity embedded in an altermagnetic superconductor. Red and blue colors denote up- and down-spin directions, that show a different spatial profile, aligned with the crystal axes.
(b) Spin-split Fermi surfaces of the altermagnetic metal considered. 
We fixed $t=1$, $E_\mathrm{F}=0.5$, $\eta_\mathrm{am}=0.25$, and $t_\mathrm{so} = 0.04t$. 
(c) Density of states for a clean altermagnetic superconductor. The system shows a hard gap between $\pm(\Delta_0 - h_0)$ where $h_0 = \eta_\mathrm{am} E_\mathrm{F}$ is the effective strength of the altermagnetism at the Fermi level. A peak of the density of states is present at $\pm(\Delta_0 + h_0$).
(d) Subgap states for a spherical and non-spherical potential impurity. In an isotropic potential, doublets of spin-degenerate subgap states appear at the impurity position. If the potential breaks the $\mathcal{C}_4$ symmetry, the spin-degeneracy is lifted and the two states acquire a polarization in the direction of $\bm{n}$, right levels.}
\end{figure}

The magnitude of the exchange field near the $\Gamma$ point depends quadratically on the electron's momentum, while the direction of the field reverses when rotating by $\pi/4$, [Fig.~\ref{fig1}(b)]. The effect of altermagnetism can be easily interpreted in the wide-band limit, $4 t\gg\mu$. We can linearize the exchange field near the Fermi level as $h_{\bm{k}} \simeq h_d(k) \cos(2 \phi_{\bm{k}})$, where $\phi_{\bm{k}}$ is the azimuth and $h_d(k) \simeq h_0 + h_1 (k-k_\mathrm{F})$, where we find $h_0 =\eta_\mathrm{am}\,E_\mathrm{F}$ as the effective exchange field strength at the Fermi level and $h_1=\eta_\mathrm{am} v_\mathrm{F}$ as its linear correction.

This work focuses on the gapped regime, $\Delta_0 > h_0$, for which the $s$-wave superconducting order parameter is stable~\cite{Chourasia_2024_Thermodynamic, Hong_2024_Unconventional}. The density of states of the pristine material $\rho(\omega) = - \pi^{-1} \Im \tr [\mathcal{G}^R(k, \omega) \sigma_0 \tau_0]$ can be calculated by integrating the retarded Green's function $\mathcal{G}^R(\omega) = (\omega\tau_0\sigma_0 -\mathcal{H}_\mathrm{b})^{-1}$ over the Brillouin zone (see Appendix~\ref{app:continuum}). The density of states features a gap for $|\omega| < \Delta_0 - h_0$ and a peak at $\omega = \pm(\Delta_0 +  h_0)$, shown in Fig.\ref{fig1}(c). This peculiar shape of the density of states, resembling a mixed $s$+$d$  superconductor~\cite{Li_1993_Mixed}, is an indicator of the presence of altermagnetism in the superconducting system that can be measured by tunneling spectroscopy. Moreover, the energy difference between the gap edge and the spectral peak allows $h_0$ to be directly inferred from tunneling spectra, providing an experimental measure of the effective exchange field.

In the following, we set $t=5$, $t_\mathrm{am} = 0.25 t$, $E_\mathrm{F}=2.5$, and $\Delta_0=1$. The main conclusions presented below do not depend on the specific choice of the parameter as long as the hierarchy of energy scales is preserved. For non-self-consistent calculation, we fix the dimension of the grid to be $81 \times 81$. We numerically diagonalize the system's Hamiltonian Eq.~\eqref{eq:H_bulk} using the \textsc{Kwant} package~\cite{Groth_2014_Kwant}, so that $\mathcal{H} |\psi_n\rangle = \omega_n | \psi_n \rangle$. We label the states by integers $n$ at $b=0$ starting from the Fermi level and using positive (negative) numbers for states above (below) $E_{\mathrm{F}}$. 

Below, we focus on the energy-resolved local density of states at the subgap states, $\rho_n (\bm{r}) = \langle \delta(\bm{r}) \tau_0 \sigma_0 \rangle_n$, and the magnetic dipole moment density along $\bm{n}$, for $\mu_n = \langle \delta(\bm{r}) \tau_0 \bm{\sigma} \cdot \bm{n} \rangle_n$, which is connected to the spin polarization of the current measured by a tunnel probe near the state.

\section{Single impurity}
Potential impurities generally do not induce subgap states in conventional superconductors~\cite{Balatsky_2006_Impurity}. This is understood as a consequence of the Anderson theorem that guarantees the robustness of the electron pairing between time-reversed states, even with disorder~\cite{Anderson_1959_Theory}. Altermagnetic superconductors break time-reversal symmetry, opening the possibility for potential impurities to generate subgap states well-separated from the continuum. The essential requirement is that the impurity locally \emph{increases} the effective exchange field, $h_d \sim k_{\mathrm{F}}^2 $, relative to the bulk. In our minimal model, where the spin splitting grows monotonically with electron density, this occurs naturally for donor-type ($V_0<0$) impurities, which raise the local chemical potential and thus the Fermi momentum. On the other hand, if the band structure exhibits a non-monotonic dispersion (for instance, by including next-to-nearest neighbor hopping so that $k_\mathrm{F}$ can shrink with increasing $\mu$), an acceptor-type ($V_0>0$) impurity could similarly enhance the local exchange splitting and generate subgap states.

Subgap states are generally always present when the energy renormalization due to the impurity is sufficiently large. When the impurity potential preserves the bulk symmetries, e.g., a rotational invariant impurity, the spin degeneracy is preserved (see Appendix ~\ref{app:extra_num}). In contrast, an arbitrary potential breaking the $\mathcal{C}_4$ or $\mathcal{M}_{xy}$ symmetries splits the spin-doublet due to the breakdown of the generalized Kramers' degeneracy, illustrated schematically in Fig.~\ref{fig1}(d). Note that other deviations from rotational symmetry effects, such as weak disorder or the presence of the sample border nearby, can also break the spin degeneracy as they mix the subgap states of the impurity.

\begin{figure}
\centering
\includegraphics{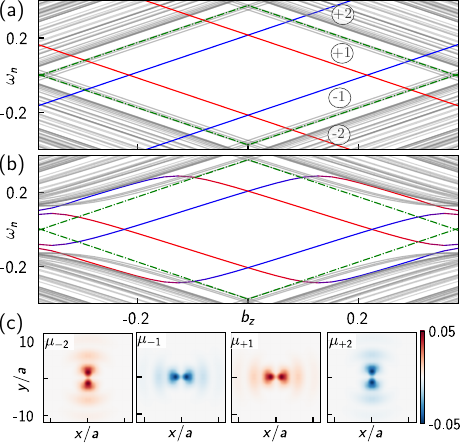}
\caption{
\label{fig2}
(a) Spectrum of an altermagnetic superconductor near a potential impurity under the action 
of an external Zeeman field. The first two subgap states ($n=+1,\ +2$) and their time-reversed twin ($n=-1,\ -2$) are colored
 red and blue depending on the magnetic dipole sign. The other states are colored in gray. The analytic gap is plotted with the green dashed line. 
(b) Spectrum for the same parameters in the presence of spin-orbit coupling, $t_\mathrm{so} = 0.02 t$. A second pair of subgap states can be seen just below the continuum. 
(c) Spin-resolved density of states of the lowest subgap states, with the sign indicating the spin direction. Parameters for the impurity: $w_\mathrm{imp} = 2.2a$ and $V_0 = -19 \Delta$. 
}
\end{figure}

A magnetic field, $\bm{b}$, parallel to $\bm{n}$ causes a Zeeman splitting of the whole spectrum and thus lifts the spin degeneracy as shown in Fig.~\ref{fig2}, where the pair of states satisfies the $\mathcal{C}_4 \mathcal{T}$ symmetry. The spectral gap closes approximately linearly following $\omega_\mathrm{mg} \approx \Delta_0 - h_0 - |b|$, shown as a dotted-dashed line. The lower state will cross the Fermi level for sufficiently strong magnetic fields [Fig.~\ref{fig2}(a)]. Unlike impurity states in conventional superconductors, a weak spin-orbit coupling does not hybridize the spin-split states [Fig.~\ref{fig2}(b)]. 

\begin{figure}
    \centering
    \includegraphics{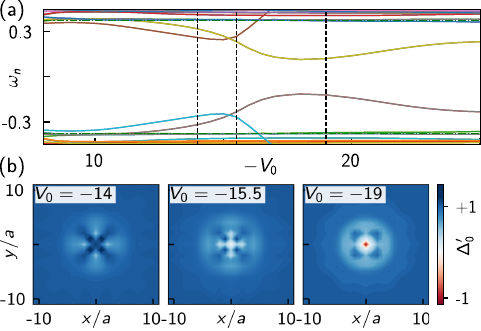}
    \caption{
    (a) Spectrum of a system with a potential impurity with increasing $V_0$. Black dashed lines mark the values used in the lower plots.
    (b) Order parameter around an impurity for three different values of $V_0$. The results in this figure have been calculated self-consistently on a $61\times61$ lattice setting $g \approx 10$ that leads to $|\Delta_0|\approx 1 $ in the bulk (further details can be found in Appendix~\ref{app:self-consistency}). 
    }
    \label{fig:3}
\end{figure}

So far, we have ignored the renormalization of the superconducting gap around the impurity, which can induce additional phenomena~\cite{Salkola_1997_Spectral, Flatte_1997_Local, Balatsky_2006_Impurity, Meng_2015_Superconducting, Mashkoori_2017_Impurity}. The effect of altermagnetism on the superconducting pairing has been explored in Refs.~\cite{Chakraborty_2024_Zero, Hong_2024_Unconventional}. For simplicity, we choose a gauge where the superconducting pairing is real. Self-consistency on the order parameter reduces the energy of the subgap states, further separating them from the continuum [Fig.~\ref{fig:3}(a)]. Moreover, the impurity renormalizes the order parameter in its vicinity, maximally suppressing it in the direction of the crystal axes [Fig.~\ref{fig:3}(b)]. 

As $V_0$ decreases, we observe a sign change in the self-consistent superconducting order parameter around the impurity. Unlike the conventional $\pi$ transition in YSR systems -- where the sign reversal is linked to a ground-state parity change triggered by a bound state crossing the Fermi level -- this transition occurs even though both subgap states remain within the gap and away from zero energy. While a parity-driven transition remains possible in our model, the sign change mechanism discussed here arises from a wavefunction-driven crossover rather than a quantum phase transition: It takes place near the value of $V_0$ where the two localized subgap states cross in energy, reflecting a reordering of spin-polarized and spatially structured wavefunctions. The transition is reminiscent of the one happening in magnetic impurities on superconductors as a function of the coupling to the substrate, transitioning from YSR-like to Andreev-like physics~\cite{Salkola_1997_Spectral, Flatte_1997_Local, Meng_2015_Superconducting, Bjoernson_2017_Superconducting}.

\section{Coupled impurities}
\begin{figure}[ht]%
\centering
\includegraphics{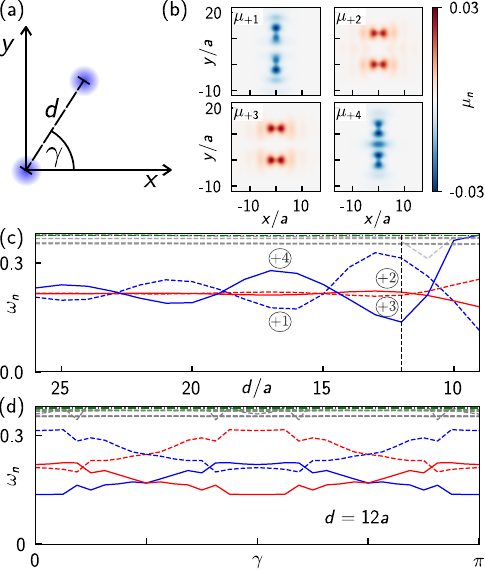}
\caption{
(a) Sketch of two coupled impurities, separated by a vector $\bm{d}$.
(b) Magnetic dipole density of states at two coupled impurities where the sign indicates the spin-up/spin-down component. The impurities are placed at distance $d = 12a$ along the $y$-axis.
(c) Subgap energy spectrum of two coupled impurities, with blue/red indicating up/down spin. The coupling favors the lowest-energy state to have parallel magnetization with either a bonding (solid line) or an anti-bonding (dashed line) distribution, depending on the distance between the impurities. 
(d) Energy of the subgap states as a function of the angle $\gamma$ between the impurities for $d \simeq 12a$. The positions of the impurities are approximated to the closest sites.
\label{fig4}
}
\end{figure}

Coupled impurities have been one of the most promising platforms to study new quantum effects, including topological superconductivity~\cite{Pawlak_review2019, Ding_2021_Tuning}. In altermagnetic superconductors, the interaction between the subgap states formed around two impurities depends on the separation vector between them, $\bm{d} = d (\cos\,\gamma,\, \sin\,\gamma\,)$, where $d$ is the distance between the impurities and $\gamma$ is the angle with the crystal axes [Fig.~\ref{fig4}(a)]. The orthogonal distribution of the two subgap spin components tends to couple the subgap states oriented parallel to the $\bm{d}$ vector [Fig.~\ref{fig4}(b)]. It leads to an energy splitting of two spin components while keeping the other two almost degenerate [Fig.~\ref{fig4}(c)]. The exchange coupling favors either the bonding or anti-bonding configurations depending on the distance with pair oscillations with an approximate period given by the Fermi wavelength of the system [Fig~\ref{fig4}(c)].
The degeneracy is lifted for any angle between impurities, except for $\gamma=j\pi/4$, $j$ being integer, where the bulk symmetries are preserved, Fig.~\ref{fig4}(d).

\begin{figure}[ht]%
\centering
\includegraphics{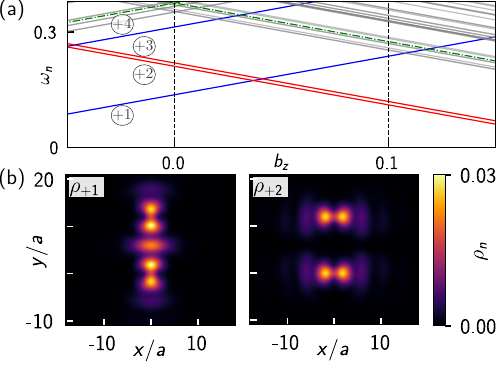}
\caption{(a) Subgap spectrum of a system with two impurities at a distance $\bm{d} = 12 \bm{e}_y$ under an externally applied magnetic field $b_z$. 
(b) Local density of states at the energy of the lowest subgap state for $b_z=0$ and $0.1$, showing the transition between two coupled to disconnected impurity states.
}\label{fig5}
\end{figure}

The lowest-energy state of the two coupled impurities can be controlled using an external magnetic field parallel to $\bm{n}$. As a function of $\bm{b}$, there is a transition from a lowest-energy subgap state that connects the two impurities to two almost-degenerate states that do not connect the impurities [Fig.~\ref{fig5}]. The transition is accompanied by a change in the spin of the lowest-energy subgap state. This transition can be measured using local spectroscopic methods, such as scanning tunneling microscopy, and allows one to switch on and off the coupling between subgap states in superconductors.

\section{Experimental implementations}
Native altermagnetic superconductivity has been predicted in some materials, such as $\mathrm{La}_2\mathrm{CuO}_4$~\cite{Smejkal_2022_Conventional, Brekke_2023_Two}. In addition, superconductivity and altermagnetism can coexist in hybrid heterostructures. The predicted signatures, i.e., directional spin-polarized subgap states, spin-split states in coupled impurities, and tunability with the external magnetic field, can be observed in both situations. These features originate from the finite-momentum properties of the altermagnet, which are projected to zero momentum due to the electron scattering with the impurity. The spatial extension of the states can be measured using scanning tunneling microscopy close to the impurity~\cite{Ruby_2016_Orbital}, while the magnetic field provides control of the energy of the states. The exotic spatial dependence of the lowest-energy subgap states with a magnetic field would constitute direct proof of the coexistence between altermagnetism and superconductivity.

\section{Conclusions}
Altermagnetic superconductors have been predicted in the literature but not experimentally demonstrated so far. In these materials, a potential impurity induces a pair of spin-polarized subgap states that extend in different crystallographic directions that are either spin degenerate or spin split, depending on whether the impurity preserves the crystal symmetries. The spatial extension of the subgap states, a characteristic signature of altermagnetism in the system, can be measured using scanning tunneling microscopy. A magnetic field parallel to the Néel vector can be used to tune the doublet splitting, whose components align in different crystallographic directions, lowering the symmetry of the local density of states. The combination of these signatures is unique to altermagnetic superconductors and their measurement, using, for example, local spectroscopic measurements, would constitute a conclusive demonstration of the coexistence of these two orders.

Devices for quantum information processing~\cite{Mishra_2021_Yu, Steffensen_2024_YSR, Geier_2024_Fermion} and topological superconductivity~\cite{NadjPerge_2014_Observation} require tuning the coupling between superconducting subgap states in a controlled manner. We have shown that the peculiar spatial extension leads to a spin-dependent coupling between impurities dependent on their orientation with respect to the crystal axes.
Moreover, the coupling between altermagnetic superconductor subgap states can be controlled using an external magnetic field, providing a knob for \textit{in situ} control of devices. The peculiar coupling between impurities can be used to explore quantum phenomena such as magnetic frustration at the atomic scale and engineer different kinds of coupling between impurities, opening the doors to different engineered phases of matter.

\textit{Note added.} Recently, we became aware of an independent
work reporting similar findings was published~\cite{Gondolf_2025_Local}.

\section{Data Availability Statement}
The data and simulation code that support the findings of
this article are available at Ref.~\cite{ZR}.
 
\section{Acknowledgments}
We acknowledge useful discussions with Adrien Bouhon and Annica Black-Schaffer. A.M. acknowledges funding from the Wallenberg Initiative on Networks and Quantum Information (WINQ). RSS acknowledges financial support from the Spanish Comunidad de Madrid (CM) ``Talento Program'' (Project No. 2022-T1/IND-24070) and the Spanish Ministry of Science, Innovation, and Universities through Grant No. PID2022-140552NA-I00.

\appendix
\section{Bulk spectrum}
\label{app:continuum}
In this appendix, we consider the bulk properties of a pristine system, i.e., a perfect lattice without impurities. 
The Hamiltonian in Eq.~\eqref{eq:H_bulk} can be expressed in momentum space as the Bloch Hamiltonian,
\begin{equation}
\label{eq:Bloch_Hamiltonian}
\begin{split}
    \mathcal{H} = & 2 t \left( 2 - \cos(a k_x) - \cos(a k_y) - E_\mathrm{F} \right) \sigma_0\tau_z \\
    &+ t_\mathrm{so} (\sin(a k_x) \sigma_y - \sin(a k_x) \sigma_x) \tau_z \\
    &+ \left[2 \eta_\mathrm{am} t \left(\cos(a k_x) - \cos(a k_y) \right) \bm{n} + \bm{b}\right] \cdot \bm{\sigma} \tau_0 \\
    &+ \Delta_{0} (\cos(\theta) \tau_x - \sin(\theta) \tau_y) \sigma_0\,.
\end{split}
\end{equation}
where $(k_x, k_y)$ is the crystal momentum, $a$ is the unit cell, $\sigma_i$ and $\tau_i$ are the Pauli matrices in the spin and particle-hole space, respectively,  $t$ is the normal hopping amplitude, $E_\mathrm{F}$ is the Fermi energy, $t_\mathrm{so}$ is the spin-orbit coupling amplitude, $\eta_\mathrm{am}$ is the altermagnet relative strength, $\bm{n}$ is the Néel vector, $\bm{b}$ is the external magnetic field, $\Delta_0$ is the superconducting order parameter amplitude, and $\theta$ is the superconducting phase.

When the $E_\mathrm{F} \ll 2t$, the Fermi surfaces are close to the $\Gamma$ point, we can work in the long-wavelength limit, $a k \ll 1$, and Taylor expand Eq.~\eqref{eq:Bloch_Hamiltonian}  as
\begin{equation}
\label{eq:continuum_bdg}
\begin{split}
    \mathcal{H} = &\left( \frac{k^2}{2m} - E_\mathrm{F} \right) \sigma_0 \tau_z \\
    + & \alpha_z k (\cos(\phi_{\bm{k}}) \sigma_y - \sin(\phi_{\bm{k}}) \sigma_x) \tau_z \\
    + &\left[ \frac{\eta_\mathrm{am} k^2}{2 m} \cos(2\phi_{\bm{k}}) \bm{n} + \bm{b} \right] \cdot \bm{\sigma} \tau_0 \\
    + &\Delta_{0} \left[ \cos(\theta) \tau_x - \sin(\theta) \tau_y \right] \sigma_0\,,
\end{split}
\end{equation}
where we defined the azimuth $\phi_{\bm{k}} = \arctan(k_y / k_x)$, the effective mass $m = 1/(2 t a^2)$, and the Rashba coupling $\alpha_z = a t_\mathrm{so}$.

When $\alpha_z = 0$, $\bm{b} = 0$, and $\theta = 0$, the spin $\sigma$ is a good quantum number, and the retarded Green's function of the bulk Hamiltonian takes a convenient form,
\begin{equation}
\label{eq:gf_complete}
    \mathcal{G}^R_\sigma(\omega, \bm{k}) = \frac{(\omega - \sigma h_{\bm{k}})\tau_0 + \xi_{\bm{k}} \tau_z + \Delta_0 \tau_x }{(\omega - \sigma h_{\bm{k}})^2 - \Delta_0^2 - \xi_{\bm{k}}^2}\,,
\end{equation}
where $\xi_{\bm{k}}$ represents the time-reversal symmetric part of the electron Hamiltonian (metal-state dispersion), and $h_{\bm{k}}$ is the time-reversal symmetry-breaking part (effective exchange field).
 
Assuming $2t \gg E_\mathrm{F} \gg 0$, we proceed by linearizing around the unperturbed Fermi surface,
\begin{equation}
    k_\mathrm{F} \approx \sqrt{2 m E_\mathrm{F}}\,,
\end{equation}
to obtain the normal state dispersion relation
\begin{equation}
    \xi_{\bm{k}} = \frac{k^2}{2 m} - E_\mathrm{F} \approx v_\mathrm{F} q \,,
\end{equation}
with $v_\mathrm{F} = k_\mathrm{F} /m$, and similarly the effective exchange field
\begin{equation}
    h_{\bm{k}} = \frac{k^2}{2 m} \eta_\mathrm{am} \cos(2 \phi_{\bm{k}}) \approx (h_0 + h_1 q) \cos(2 \phi_{\bm{k}})\,,
\end{equation}
with $h_0 = \eta_\mathrm{am} E_\mathrm{F}$ and $h_1 = \eta_\mathrm{am} v_\mathrm{F}$, and $q = k - k_\mathrm{F}$.

The size of the spectral gap can be found by considering the direction with the strongest effective exchange field, $\phi_{\bm{k}} = n\pi/2$, resulting in:
\begin{equation}
    \omega_\mathrm{mg} 
    \approx \Delta_{0} - \eta_\mathrm{am} E_\mathrm{F}\,.
\end{equation}
Therefore the system becomes gapless, $\omega_\mathrm{mg}=0$, when $\Delta_{0} \sim E_\mathrm{F}$, indicating that even a very small altermagnetism is enough to close the gap when the Fermi energy is large, justifying our focus on the long-wavelength approximation regime of validity. Note that in the direction $\phi_{\bm{k}} = \pm \pi/4, \pm 3\pi/4$, the gap remains open regardless of the strength of altermagnetism, with a directional spectral gap equal to $\Delta_{0}$. These considerations are valid as long as the gap is induced (i.e., we are not considering self-consistent solutions).

\begin{widetext}
We obtain the density of states of the clean bulk material by integrating the Green's function over the first Brillouin zone
\begin{equation}
\begin{split}
    g^R_\sigma(\omega) &= \frac{L^2}{4\pi^2} \iint_{BZ} \dd k_x\ \dd k_y\ \mathcal{G}^R_\sigma(\omega, \bm{k}) \\
    &\approx \frac{L^2}{4\pi^2} \int_0^{2\pi} \dd \phi_{\bm{k}} \int_0^{+\infty} k\ \dd k\ \mathcal{G}^R_\sigma(\omega, \bm{k}) \\
    &\approx \frac{L^2}{4\pi^2} \int_0^{2\pi} \dd \phi_{\bm{k}} \int_{-q_\mathrm{c}}^{+q_\mathrm{c}} \dd q\ (k_\mathrm{F} + q) \frac{\left[\omega - \sigma \left[h_0 + h_1 q \right] \cos(2 \phi_{\bm{k}}) \right]\tau_0 + v_\mathrm{F} q \tau_z}{\left(\omega - \sigma \left[h_0 + h_1 q \right] \cos(2 \phi_{\bm{k}})\right)^2 - \Delta_{0}^2 - \left(v_\mathrm{F} q \right)^2} \\
\end{split}
\end{equation}
where in the third line we linearized around the Fermi momentum $k_\mathrm{F}$ and integrated on a symmetric region $[-q_c, +q_c]$. To handle the integral, we note that there are terms in the numerator of order 0, 1, and 2 in $q$. The first-order terms vanish because of the symmetry of the integration region, while the term proportional to $q^2$ is linearly divergent in $q_c$, signaling the breakdown of the linear approximation.

Keeping only the zeroth-order term we get 
\begin{equation}
\begin{split}
   &\frac{L^2 k_\mathrm{F}}{4\pi^2} \int_0^{2\pi} \dd \phi_{\bm{k}} \int_{-q_\mathrm{c}}^{+q_\mathrm{c}} \dd q\ \frac{\left[\omega - \sigma \left[h_0 + h_1 q \right] \cos(2 \phi_{\bm{k}}) \right]\tau_0}{\left(\omega - \sigma \left[h_0 + h_1 q \right] \cos(2 \phi_{\bm{k}})\right)^2 - \Delta_{0}^2 - \left(v_\mathrm{F} q \right)^2}  \\
    &\approx -\frac{L^2 k_\mathrm{F}}{4 \pi^2} \frac{2}{v_\mathrm{F}} \int_0^{2\pi} \dd \phi_{\bm{k}} \frac{\left[\omega - \sigma h_0 \cos(2 \phi_{\bm{k}})\right] \tau_0 + \Delta_{0} \tau_x}{\sqrt{\left[\omega - \sigma h_0 \cos(2 \phi_{\bm{k}})\right]^2 -  \Delta_{0}^2 \left[1 - \eta_\mathrm{am}^2 \cos^2(2 \phi_{\bm{k}})\right]}}  \times M(\omega).
\end{split}
\end{equation}
with the factor
\begin{equation}
   M(\omega) = - \mathrm{arctanh} \left(\frac{v_\mathrm{F} q_c}{\sqrt{\left[\omega - \sigma h_0 \cos(2 \phi_{\bm{k}})\right]^2 -  \Delta_{0}^2 \left[1 - \eta_\mathrm{am}^2 \cos^2(2 \phi_{\bm{k}})\right]}}\right) .
\end{equation}
which, for large $q_c$, converges to $M(\omega) \to i \pi /2$. By considering a large cutoff energy $v_\mathrm{F} q_c$ with respect to $\Delta_{0}$, the final result is then
\begin{equation}
\label{eq:gf_approx}
\begin{split}
   g^R_\sigma(\omega) \approx -  i \pi \frac{\rho_0}{2} \frac{1}{2\pi}  \int_0^{2\pi} \dd \phi_{\bm{k}} \frac{\left[\omega - \sigma h_0 \cos(2 \phi_{\bm{k}})\right] \tau_0 + \Delta_{0} \tau_x}{\sqrt{\left[1 - \eta_\mathrm{am}^2 \cos^2(2 \phi_{\bm{k}})\right] \Delta_{0}^2 - \left[\omega - \sigma h_0 \cos(2 \phi_{\bm{k}})\right]^2}}\,,
\end{split}
\end{equation}
where $\rho_0 = {m L^2}/{\pi}$ is the normal-state density of states and the final integral on the azimuthal angle must be performed numerically. Then the density of states is calculated as $\rho(\omega) = - \pi^{-1} \mathrm{Im}\ \tr_{\tau\sigma} g^R_{\tau \sigma} (\omega)$.
\end{widetext}

\begin{figure}[h]
    \centering
    \includegraphics[]{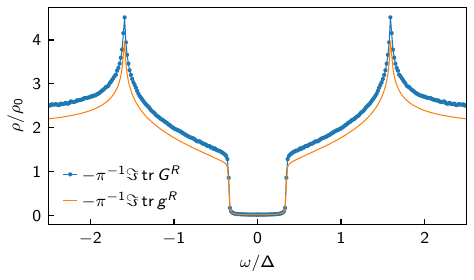}
    \caption{Comparison of the semiclassical approximation for the density of states of an altermagnetic superconductor with the exact result calculated by direct integration of $G^R$ over the Brillouin zone.
    Parameters: $t=5$, $\mu=2.5$, $\eta=0.25$, $\Delta_{0}=1$}.
    \label{fig:dos_comparison}
\end{figure}

We compare the result to a direct numerical integration of Eq.~\eqref{eq:gf_complete} with Eq.~\eqref{eq:gf_approx} and show it in Fig.~\ref{fig:dos_comparison}. 

\section{Self-consistent order parameter}
\label{app:self-consistency}
In this appendix, we examine the details of the self-consistent calculation of the order parameter. We recall that the relationship between the Green's functions and the BdG Hamiltonian can be expressed as follows,
\begin{equation}
    (\omega - \mathcal{H})\ \mathcal{G} =
    \begin{pmatrix}
        \omega - \mathcal{H}_0 & \Delta \\
        \Delta^\dag & \omega + \mathcal{T} \mathcal{H}_0 \mathcal{T}^{-1} 
    \end{pmatrix}
    \begin{pmatrix}
        G & F \\
        F^\dag & \tilde{G} 
    \end{pmatrix} = \mathbb{I}\,,
\end{equation}
where $\mathcal{G}$ is the Nambu Green's function, while $G$ and $F$ are the normal and anomalous Green's functions, respectively. We can express the Green's function in terms of the eigenstates as
\begin{equation}
    \mathcal{G}(\omega) = (\omega - \mathcal{H})^{-1} = \sum_s \psi_s^\dag \psi_s \delta(\omega - \omega_s).
\end{equation}
In particular, the real and imaginary parts of the anomalous correlations are
\begin{align}
    F_\mathrm{re}(\omega) &= \frac{1}{2} \tr[\tau_x \mathcal{G}] = \frac{1}{2} \sum_s \langle \tau_x \rangle_s \delta(\omega - \omega_s), \\
    F_\mathrm{im}(\omega) &= \frac{-i}{2} \tr[\tau_y \mathcal{G}] = \frac{-i}{2} \sum_s \langle \tau_y \rangle_s \delta(\omega - \omega_s).
\end{align}
The gap equation can then be written in terms of the expectation value of the anomalous correlator for a thermal state,,
\begin{equation}
    \Delta_0 
    = g \int_{-\infty}^{+\infty} \dd\omega\, F_0(\omega) \tanh\left(\frac{\omega}{2T}\right),
    \label{eq:gap_equation}
\end{equation}
where $F_0$ is the singlet component. The magnitude and phase of the order parameter can be obtained as
$|\Delta_0| = g \sqrt{\langle \tau_x \rangle^2 + \langle \tau_y \rangle^2}$ and
$\theta = \arctan(\langle \tau_y \rangle / \langle \tau_x \rangle)$, where the expectation values are taken against the distribution $\tanh(\omega / 2T)$.

In a bulk system, Eq.~\eqref{eq:gap_equation} is generically divergent. In a numerical implementation, the discretization on a finite-size lattice automatically renormalizes the UV divergence by introducing an upper cutoff in the momentum and consequently in the energy. However, computing all the eigenstates of the lattice Hamiltonian is computationally expensive. We can address this problem and approach the standard BCS treatment by introducing an energy cutoff $\omega_\mathrm{c}$ such that 
\begin{equation}
    \Delta_0 = g \int_{-\omega_\mathrm{c}}^{+\omega_\mathrm{c}} \dd\omega\, F_0(\omega) \tanh\left(\frac{\omega}{2T}\right).
\end{equation}

The model then depends on the specific choice of $g$ and $\omega_\mathrm{c}$. We avoid this issue by considering that the physically meaningful quantity is only $\Delta_0$ and proceed as follows. We fix $\omega_\mathrm{c}$ implicitly by calculating only a sufficiently high number of eigenstates $N_\mathrm{c}$ such that the maximum value of $\omega_n$ is higher than the peak position $\omega_\mathrm{p} = \Delta_0 + \eta E_\mathrm{F}$. We then compute the unnormalized gap
\begin{equation}
    \tilde{\Delta}_0 = \frac{1}{2} \sum_{n=-N_\mathrm{c}/2}^{N_\mathrm{c}/2} \left[\langle \tau_x \rangle_n - i \langle \tau_y \rangle_n \right] \tanh\left(\frac{\omega_n}{2T}\right).
\end{equation}
Finally, we set $g$ equal to the inverse of the average unnormalized gap:
\begin{equation}
    g \simeq \frac{\Omega}{\int_\Omega \dd \bm{r} |\tilde{\Delta_0}(\bm{r})|}.
\end{equation}
In this way, the order parameter $|\Delta_0|$ is approximately always equal to 1 far from the impurity. Using a grid of $61 \times 61$ and a cutoff of $N_\mathrm{c} = 1024$, we find that $g \approx 10$. Once $N_\mathrm{c}$ and $g$ are set, we can proceed with a standard self-consistent loop until convergence.

\section{Additional numerical results}
\label{app:extra_num}
\begin{figure}[ht]
    \centering
    \includegraphics{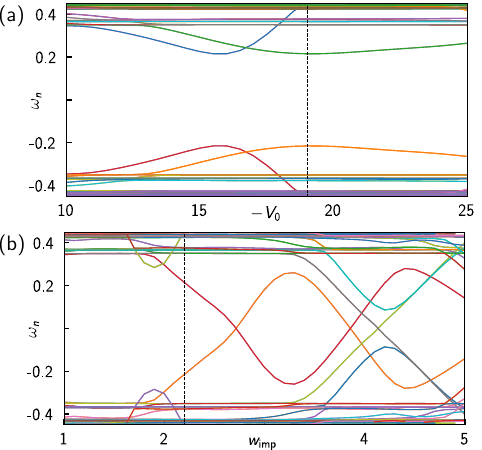}
    \caption{Low-energy spectrum of an altermagnetic superconductor with different Gaussian impurities.
    (a) shows the bound states energy for $w_\mathrm{imp} = 2.2 a$, where the first one and then a second pair of spin-degenerate bound states separates from the continuum as the strength of the potential is increased.
    (b) We fix $V_0 = - 19 \Delta_0$ and increase the impurity size. Around $w_\mathrm{imp}=2$ a couple of spin-degenerate pairs separate from the spectrum and oscillate. With increasing size, there is a crossing of the Fermi energy and then another doublet separate from the continuum.
    }
    \label{fig:S1}
\end{figure}

In this appendix, we show some additional results for different values of the impurity potential parameters. The parameters used are the same as in the main text to allow direct comparison, i.e., $t = 5.0$, $\eta_\mathrm{am} = 0.25$, $\mu = 2.5$. In Fig.~\ref{fig:S1} we show the low-energy spectrum of the one-impurity system for various $V_0$ and $w_\mathrm{imp}$ obtained by exact diagonalization of the tight-binding Hamiltonian on an $81\times81$ site system.
Depending on the size and depth of the potential created by the impurity, additional pairs of subgap states can appear. Furthermore, increasing $V_0$ increases the gap between the states and the continuum, bringing them toward the superconductor's Fermi level; some of them cross the Fermi level. Higher states will feature nodes in the quasiparticle wavefunction, changing the interaction between subgap states.

\begin{figure}[hb]
    \centering
    \includegraphics{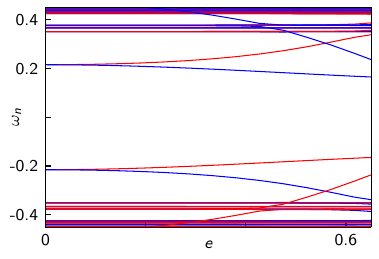}
    \caption{Lifting of the spin degeneracy of a pair of subgap states due to increased eccentricity of the impurity potential.
    A finite eccentricity $e$ lifts the degeneracy of the spectrum and splits the spin doublets. Each eigenstate has a definite spin quantum number, which we show in the color of the line. }
    \label{fig:S2}
\end{figure}

We also studied a generalized Gaussian impurity defined as 
\begin{equation}
    f_i(\bm{r}) = {\exp\left[  \frac{ -(\bm{r} - \bm{r}_i) \cdot A_i (\bm{r} - \bm{r}_i)}{2} \right]}\,,
\end{equation}
where the matrix $A$ is defined by 
\begin{equation}
     A_i^{-1} = R_z(\gamma_i) \frac{w_\mathrm{imp}^2}{2}  \begin{pmatrix} \sqrt{1+ {e}_i^2} & 0 \\ 0 & \sqrt{1+ {e}_i^2} \end{pmatrix} R_z(\gamma_i)^T \,,
\end{equation}
where $\sigma$ parametrizes the size, $e$ the eccentricity, and $\alpha$ the rotation with respect to the crystal axes. In Fig.~\ref{fig:S2} we show the spin-polarizing effect of eccentricity. This result motivates the sketch in Fig.~\ref{fig1}(d).

\clearpage
\bibliography{am_impurities}
\end{document}